\documentclass[preprint,showpacs,preprintnumbers,amsmath,amssymb]{revtex4}

\usepackage{graphicx}
\usepackage{dcolumn}
\usepackage{bm}

\begin{document}

\title{Exceptionally Slow Rise in Differential Reflectivity Spectra of
Excitons in GaN: Effect of Excitation-induced Dephasing }

\author{Y. D. Jho}
\altaffiliation[Current Address: ]{National High Magnetic Field
Laboratory, Florida State University, Tallahassee, FL 32310}
\email{ydjho@magnet.fsu.edu}
\author{D. S. Kim}
 \affiliation{Department of Physics,
Seoul National University, Seoul 151-747, Korea}
\author{A. J. Fischer and J. J. Song}
\affiliation{Department of Physics and Center for Laser Research,
Oklahoma State University, Stillwater, Oklahoma 74078}
\author{J. Kenrow}
\altaffiliation[Current Address: ]{Dept. of Electrical and
Computer Engineering, University of the Pacific, Stockton, CA
95211}
\author{K. El Sayed}
\author{C. J. Stanton}
\affiliation{Department of Physics, University of Florida,
Gainesville, Florida 32611}

\begin{abstract}

Femtosecond  pump-probe (PP) differential reflectivity spectroscopy
(DRS) and four-wave mixing (FWM) experiments were performed
simultaneously to study the initial temporal dynamics of the exciton
line-shapes in GaN epilayers.  Beats between the A-B excitons were
found \textit{only for positive time delay} in both PP and FWM
experiments. The rise time at negative time delay for the differential
reflection spectra was much slower than the FWM signal or PP
differential transmission spectroscopy (DTS) at the exciton resonance.
A numerical solution of a six band semiconductor Bloch equation model
including nonlinearities at the Hartree-Fock level shows that this slow
rise in the DRS results from excitation induced dephasing (EID), that
is, the strong density dependence of the dephasing time which changes
with the laser excitation energy.

\end{abstract}

 \maketitle

\section{Introduction}

Group III nitride semiconductors such as GaN and InGaN have become
important materials owing to their optoelectronic device
applications in the blue and ultraviolet spectral range and their
use in high temperature electronic devices. The demonstration of
InGaN multiple quantum well laser diodes\cite{Nakamura} has also
inspired tremendous research interest in the nitride-based
materials. Transient four-wave mixing (FWM) studies on GaN were
performed to investigate the intrinsic excitonic
properties\cite{Fischer,Pau} and the influence of electron spins
on exciton-exciton interaction\cite{Gonokami}. Femtosecond
pump-probe (P-P) measurements were done by C.-K. Sun \textit{et
al.} on InGaN\cite{Sun1} and n-doped GaN\cite{Sun2} to investigate
the fast carrier cooling. Time-resolved studies of coherent
acoustic phonons in GaN and GaN/InGaN systems
\cite{Sun99,Sun2000,Yahng1,Stanton2002}, as well as coherent
optical phonons\cite{Yee2002} were performed. Recently, field
dependent carrier decay dynamics were done by Jho \textit{et
al.}\cite{Jho2001, Jho2002}

In the FWM experiments on GaN, quantum beats of excitons and their
phase changes via polarization configurations have been studied
and exciton-phonon interaction rates were deduced.  Nevertheless,
there are still not a lot of  time-domain studies regarding the
coherent response of excitons in GaN including many body effects.

The FWM line shapes not only discriminate between homogeneous
broadened and inhomogeneous broadened systems\cite{Fischer} but
also provide information on the carrier-carrier
interaction\cite{Shah}. For instance, time-integrated (TI) FWM
signals at negative time delays have been observed in GaAs quantum
wells\cite{Wegener,Leo} and understood by local-field
effects\cite{Wegener,Leo} or excitation-induced depasing
(EID)\cite{Wang} which are incorporated into the semiconductor
Bloch equations. In addition, it was argued that EID dominates at
a moderately low exciton density ($<$ $10^{16} cm^{-3}$) and gives
a strong contribution to TI-FWM in the co-linearly polarized
geometry.\cite{Wang}

In this work we report the comparative studies of femtosecond
degenerate pump-probe (PP) differential reflectivity spectroscopy
(DRS) measurements and time-integrated (TI)-FWM experiments on GaN
epilayers as a function of excitation energy at low carrier
excitation density.  In addition, we have also performed PP
differential transmission spectroscopy on a thin GaN sample,
though not simultaneously with FWM. Our results show that the
quantum beats as revealed in previous
studies\cite{Fischer,Pau,Gonokami} are the same both in PP and FWM
and exist \textit{only for positive time delay}. Results further
show that the DRS and FWM differ at negative time delays, with the
DRS signal persisting longer at the negative delays in spite of
inhomogenous broadening.  Calculations based on a six-band
semiconductor Bloch equation model solved in the Hartree-Fock
level show that this difference arises from EID, that is, the
dephasing time depends on the carrier density excited by the laser
pulse.

\section{Experimental scheme}

\begin{figure}[tbp]
\includegraphics[scale=0.4]{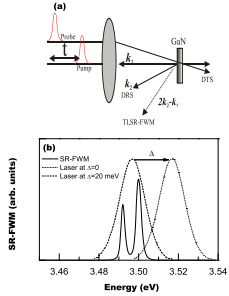}
\caption{(a) Experimental schematic showing the simultaneous
measurement of the four-wave
mixing (FWM) and  pump-probe (PP) differential reflectivity spectra (DRS).
In the reflection geometry, the wave vector of the pump ($\mathbf{k_1}$)
and probe ($\mathbf{k_2}$) are shown.
(b) The  spectrally resolved (SR)-FWM data at 11 K (solid line) for
t=0 (i.e. no time delay between pump and probe).
The line-width of
the peaks are 2.1 meV (A exciton) and 2.5 meV (B exciton). The
dashed and dotted line show the laser spectrum with detuning
$\Delta = 0 $ (and the center between the two excitons)
and 20 meV respectively. \label{fig1}}
\end{figure}

The GaN samples used in this work were a 7.2 $\mu$m-thick and a 2
$\mu$m-thick epilayer grown with the wurtzite structure on a
(0001) sapphire substrates by metalorganic chemical vapor
deposition. The second harmonic of a femtosecond Ti-sapphire laser
in the high-energy region of the tuning (705 nm - 710 nm) was
used, with 150-fs pulse-width. As shown in the Fig. \ref{fig1}(a),
we have performed both FWM and differential reflectivity
spectroscopy (DRS) {\it simultaneously} in the reflection geometry
on 7.2 $\mu$m-thick sample. To compare the differential reflection
spectra with the differential transmission spectra (DTS), a 2
$\mu$m-thick sample was also used. The pump and probe pulses were
at the same wavelengths and collinear polarization. All
measurements were performed at 11 K unless otherwise noted (c.f.
Fig.~\ref{fig7}). To reduce the effect of the laser noise, we used
a differential amplification scheme after dividing the probe beam
($\mathbf{k_1}$) into two.  With this scheme, the DRS signal can
be as small as $10^{-4}$.

In Fig.\ref{fig1}(b), the spectrally resolved (SR) FWM data (solid
line) at 11 K  is shown together with the spectrum of laser for
detunings of 0 meV (dashed) and 20 meV (dotted) from the center of the
exciton peaks. There is no time delay between pump and probe.  The two
peaks of Fig. \ref{fig1} (b) correspond to the $\Gamma^{V}_9-
\Gamma^{C}_7$ exciton (A exciton transition) and the $\Gamma^{V}_7-
\Gamma^{C}_7$ exciton (B exciton transition), which are caused by
crystal field and spin-orbit coupling. The dashed and dotted lines are
the spectrum of laser at the center of the exciton resonances and at 20
meV above the resonance. The linewidth of peaks are measured to be 2.1
meV for the A exciton, and 2.5 meV for the B exciton.  The energy
difference of the two excitons is about 8 meV.

The power of the pump (probe) pulse which is in the reflection
direction of $\mathbf{k_1}$ ($\mathbf{k_2}$) was 0.5 mW (0.1 mW).
The two beams are focused onto a 100 $\mu$m spot with the
external-crossing angle of $6^o$. From the measured absorption
coefficient, we estimate the initial carrier density to be
$5\times 10^{15} cm^{-3}$ at the center of the exciton resonances
(3.497 eV)\cite{Fischer2}. There was no detectable change of decay
time in the FWM for carrier densities ranging from $10^{15}
cm^{-3}$ to $5 \times 10^{16} cm^{-3}$.

\section{Results and discussion}

\begin{figure}[tbp]
\includegraphics[scale=0.45,trim=0 20 0 0,clip]{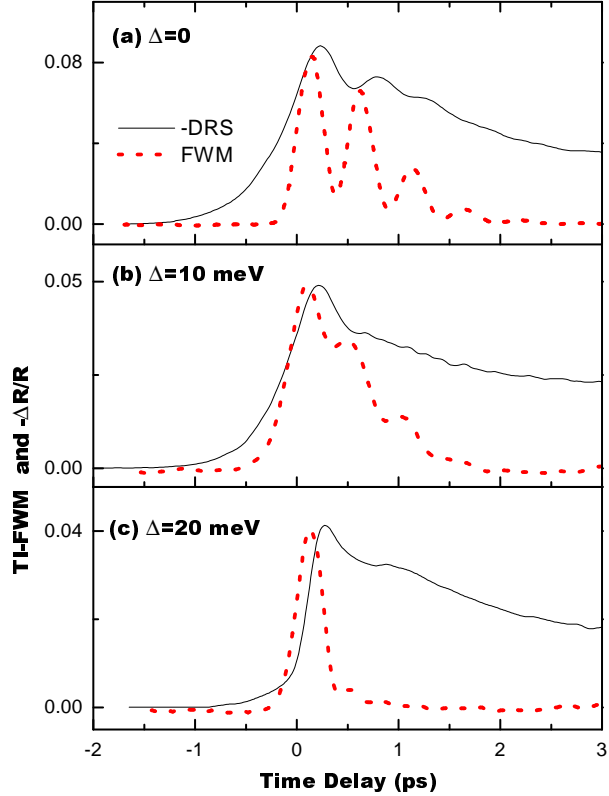} 
\caption{Experimental DRS (solid lines) and FWM (dotted lines) at
different detunings $\Delta$ above the exciton resonance energy (=
3.497 eV).  The FWM and DRS both show oscillations at positive time
delay associated with beating between the A and B excitons. At negative
time delay, the DRS shows a slow rise time while the rise time of the
FWM is determined by the pump pulse.  Exciton beating is not oberved in
negative time delay.
\label{fig2}}
\end{figure}

Fig.~\ref{fig2} shows the simultaneous measurements of the DRS and FWM
at the excitation energy of (a) 3.497 eV, (b) 3.507 eV, and (c) 3.517
eV.  The FWM data were scaled with the DRS data for comparison. In both
the PP and FWM data, the strongest signal was observed at the exciton
resonance,  3.497 eV.  This is in the middle of the A and B excitons.
Here $\Delta$, the detuning is chosen to be the energy above 3.497 eV
and hence the figures correspond to detunings of $\Delta = 0, 10$ and
$20$ meV respectively.  The PP data and the FWM, in positive time
delay, show similar features, namely that of beating  between
the A and B excitons.  This beating is strongest for $\Delta = 0$, i.e.
at the resonance and has a  period of about 500 fs. This period is
consistent with the SR-FWM that showed the energy difference of 8 meV
between A and B excitons (c.f.  Fig.~\ref{fig1}).

One puzzling difference between the PP DRS and FWM signals is the behavior
at negative time delay.  The PP signals in negative time delay persist
much longer than the FWM and do not show the A-B exciton beatings.
The DRS behavior is strongest at the exciton resonance ($\Delta=0$), but
becomes less pronounced as one increases the detuning ($\Delta=$ 20
meV) and excites further into the band.  The values of the DRS rise times
obtained from first order exponential fits were 445 fs ($\Delta = 0$),
381 fs ($\Delta$= 10 meV), and 183 fs ($\Delta$=20 meV). The error in
determining the rise time is less than 20 fs for all measurements. The
rise times of FWM were less than 200 fs for all excitation energies and
comparable to the pump laser duration.

The fact that there is a fast rise time in negative delay in the FWM is
not surprising. To see a slow rise in the  FWM signal in the negative
time delay requires that the sample be very clean and that
inhomogeneous broadening is weak.\cite {Wegener,Leo} Inhomogeneous
broadening will wash out any negative time signal in the FWM.

\begin{figure}[tbp]
\includegraphics[scale=0.45,trim=0 20 0 25,clip]{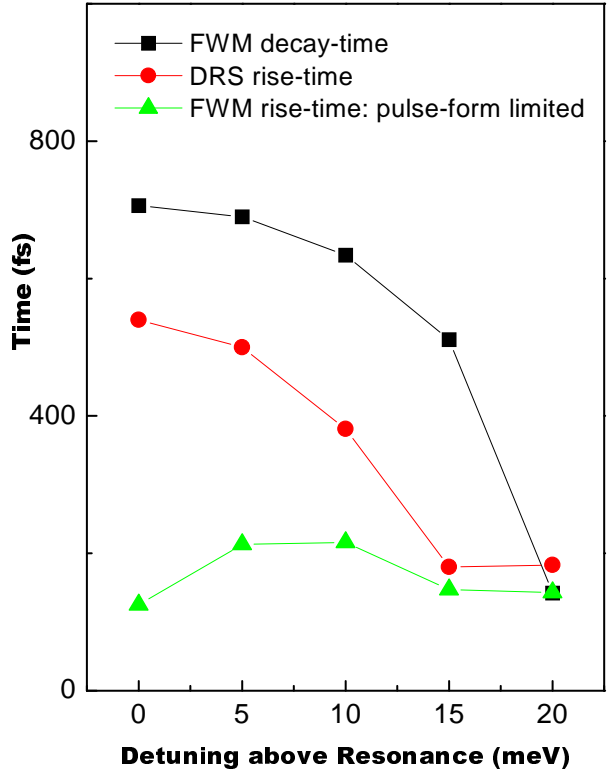}
\caption{The rise and decay times of the DRS and FWM as a function of
laser detuning above the exciton resonance.  The DRS rise time is given
by the circles, the FWM rise time by the triangles, and FWM
decay time by the squares. Note that the decay time of the FWM and the
rise time of the DRS are comparable while the rise time of the FWM appears
to be
determined by the pump laser pulse width. \label{fig3}}
\end{figure}

In Fig.~\ref{fig3} we compare the rise times of the DRS versus the FWM
signal.  Since the values of FWM decay time ($\tau_{FWM}$) for positive
delay and zero detuning ($\Delta=0$) in Fig.~\ref{fig3} are
at least three times larger than the FWM
rise times, this indicates that the spectra are inhomogeneously
broadened in these samples.\cite{Fischer,Wegener} In fact, the rise
time of the FWM signal appears to be limited by the pulse duration,
again consistent with strong inhomogeneous broadening.  As a result,
one would not expect to see a FWM signal in the negative time delay
since it has been shown that this would occur only
in a homogeneously broadened system.\cite{Wegener,Leo}

In contrast, the differential pump-probe reflectivity spectra can have
a signal at negative delay, even with inhomogeneous broadening.  This
is related to the free polarization decay (FPD) of the {\it probe}
pulse which plays a role in the DRS signal at negative time delays.
Note however, that the FPD of the probe pulse does not effect the FWM.
This is because while the FPD of the probe pulse will produce  a signal
in the probe direction ($\mathbf{k_2}$), it does not produce a signal
in the FWM direction $\mathbf{2k_2-k_1}$.\cite{Leo}

The FPD persists when the probe precedes the pump even in an
inhomogeneously broadened system.  This leads to a rise time with an
effective time constant $T_2^{\ast}$.  $T_2^{\ast}$ is given
approximately by the inverse of frequency spread due to inhomogeneous
broadening $1/\Delta\omega$ and is shorter than $T_2$, the homogeneous
dephasing time.\cite{Brito}  If the pump pulse (which is now after the
probe pulse for negative time delay) overlaps with the tail of the FPD
of the probe pulse,  then it may be possible to produce a signal in the
probe pulse.

\begin{figure}[tbp]
\begin{center}
\includegraphics[scale=.45,trim=20 0 0 40,clip]{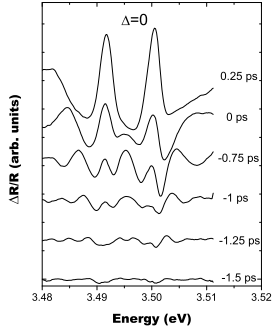}
\end{center}

\caption{Experimentally measured spectral resolved DRS.  The pump
detuning is $\Delta=0$, i.e. between the A and B exciton resonances.
The time delay between pump and probe is given for each trace
in the figure.  \label{fig4}}
\end{figure}

It has been shown in DTS, that the overlap of the pump with the tail of
the probe polarization generates a transient diffraction grating  with
wavevector $\mathbf{k_2-k_1}$.  This transient grating can lead to a
diffraction of part of the pump pulse with wave vector $\mathbf{k_1}$
into the probe direction $\mathbf{k_1+(k_2-k_1)}$.  This is the
qualitative origin of the well-known coherent oscillations in
semiconductors\cite{Koch} or the perturbed free polarization decay
term\cite{Brito}. The spectral oscillations induced from the transient
grating have no correlation with the exciton oscillations seen in
Fig.~\ref{fig2}(a) at positive time delay.  These coherent oscillations
are readily observed in our sample, and are shown in Fig.~\ref{fig4}
for the spectrally resolved DRS signal. The spectral oscillations,
depend on the time delay as well as the detuning from the exciton
resonance.

However, a solution of the density matrix equation\cite{Brito} without
incorporating the exciton-exciton interaction or EID,  shows that, when
integrated over frequency, \textit{the different spectral oscillatory
signals cancel out and the net result is that DTS spectra at each
negative time delay is integrated out to be zero.}  This result should
also hold for the DRS spectrum when the excitation is deep within the
band continuum.  In fact, Fig.~\ref{fig2}(c) precisely shows this
effect.  (In addition, for large detuning and excitation within the
band, the dephasing time $T_2$ should be much shorter than for
excitations between the A and B excitons).

Below the continuum band edge, the situation is more complex.  To
study this effect more thoroughly, calculations were performed
based on the Semiconductor Bloch Equations.

\section{Calculations}
To better understand the origin of the slow rise-time of the DRS signal
in Fig.~\ref{fig2}, we have calculated both the differential reflection
(DRS) and differential transmission spectra (DTS) by solving a coupled
six-band semiconductor Bloch equation model\cite{Kenrow,Hu} including
all Hartree-Fock nonlinearities.  From the semiconductor Bloch
equations, the dielectric response is calculated. Typically, the
reflection is much more sensitive to the real part of the dielectric
response while the transmission is more sensitive to the imaginary part
of the dielectric response. In our calculations, we have included
carrier scattering on a phenomenological level to allow for the
relaxation of the photo-excited carriers back to equilibrium. In
addition, we have included excitation induced dephasing
(EID)\cite{Sayed,Wang} to allow for the change in the carrier dephasing
time as the density of excited carriers changes with the pump laser
pulse.

In the framework of EID, the inverse polarization decay time is given
by
\begin{equation}
1/ T_2
\,  = \, 1/ T_{2,0} \,  + \, n/T_{2,1},
\label{scatter}
\end{equation}
and is a linear function of the
induced carrier density, $n$. $T_{2,1}$ corresponds to the
carrier-carrier scattering time as the carriers are photoexcited in a
nonthermalized distribution. At low carrier densities, scattering such
as electron-electron is a linear function of the density.  At higher
densities, screening effects can become important and change the
density dependence.  Note that the polarization decay time has the
important property that it is long  when the probe proceeds the pump
since no carriers have been created yet and the decay time becomes
shorter when the probe comes after the pump pulse which creates
photoexcited carriers.

We chose the low density dephasing time,
$T_{2,0}$ to be 1 ps. By numerically integrating the
semiconductor Bloch equations, we computed the induced polarization
with and without the pump pulse present. The corresponding dielectric
response, which determines the reflection and transmission spectra, is
obtained from the Fourier transform of the probe polarization.

In calculating the dielectric response, we do not allow for
\textit{intraband} changes to the dielectric function. This should be
less important when one excites below the band edge since there are no
free carriers available to screen out the laser pulse and give a
Drude-like contribution to the dielectric function, which can be
important in the reflectivity.  However, this becomes more important
when a large number of free carriers are excited above the band gap and
for strong excitation above the gap, may even dominate the
signal.\cite{Zollner1,Zollner2} In addition, we do not include
diffusion of carriers away from the
surface\cite{Zollner1,Zollner2,Bailey} of the sample in our
calculations. The intent of the calculations is to understand the
\textit{initial} behavior of the PP DRS and DTS spectra. Diffusion
effects become important on a time scale longer than 1 ps and should be
included along with more accurate scattering models in more detailed
studies for longer times.

\begin{figure}[tbp]
\includegraphics[scale=0.55,trim=10 0 0 50,clip]{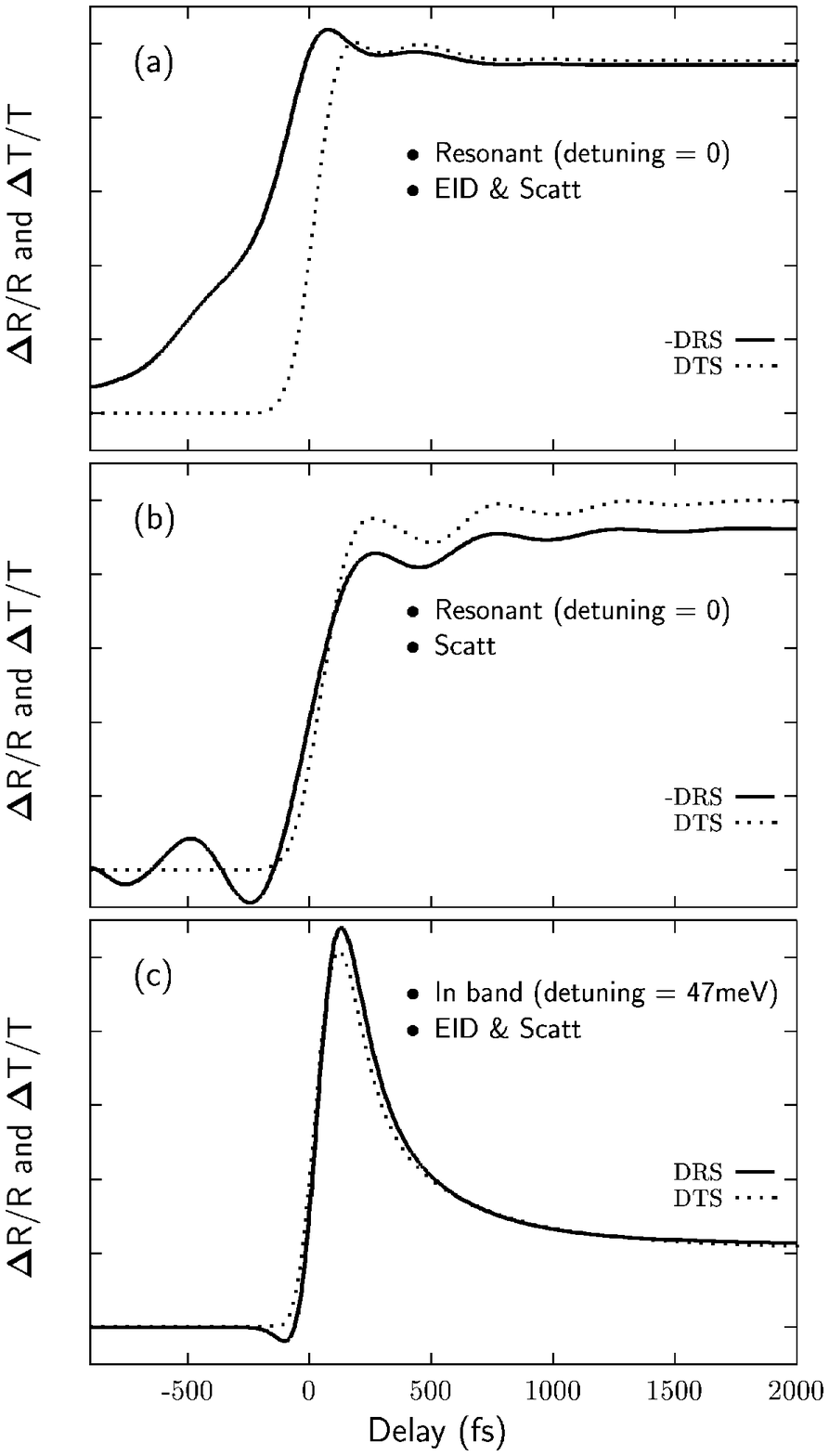}
\caption{The calculated differential reflection (DRS) and
differential transmission spectra (DTS) as a function of delay
time between the pump and probe pulses based on the Semiconductor
Bloch Equations. In (a), the laser excitation energy is resonant
between the excitons. Scattering as well as excitation induced
dephasing (EID) is included in the simulation (both terms in
eq.~\ref{scatter}.). As can be seen, the DRS signal has a slower
rise time than the DTS signal. In (b), the laser excitation energy
is resonant with the excitons but only  the first scattering term
in eq.~\ref{scatter} is included. Here both the DRS and DTS signal
have similar rise times.  In (c), the laser excitation energy is
deep into the band. Both EID and scattering are included in
simulation.  The rise time  of the DRS is faster than for laser
excitation energy at the exciton (a).\label{fig5}}
\end{figure}

Results of the numerical solution to the model are shown in figure 5.
Figure 5 shows the computed DTS and DRS signals as a function of delay
between the pump and probe pulses. Fig. 5 (a) gives the result when the
pump and probe are both at resonance with the excitons (excitation
between the A and B excitons) and EID as well as carrier scattering are
included (i.e. both terms in eq.~\ref{scatter}).  The corresponding
results when EID was excluded (i.e., only the first term on the left
hand side of eq.~\ref{scatter} is included) are shown in Fig. 5(b).
Note that both figures show the oscillations resulting from A and B
exciton beating in both the DRS and DTS signals at postive time delay.

Our simulation also reproduces the slow rise-time of the DRS signal in
Fig. 2(a) {\it only if EID is included} (cf. Fig. 5(a)). However, if
EID is excluded, the effect vanishes and the signal strength is
diminished by 60 \% (cf. Fig. 5(b).) In this case, oscillations occur
at negative time in the in the DRS. It is important to note that this
behavior is \textit{not seen} for the corresponding DTS signal where
the rise times are unaffected by EID. However, a comparison of the two
DTS signal strengths shows that the inclusion of the EID enhances the
signal strength by 40\%. When probing the samples at laser energies
deep into the band, (Fig. 5(c)) we find that the rise time of the DRS
signal is now faster in agreement with experiment (cf.  Fig. 2, 3), and
shows little difference with the DTS signal.  Of course, deep within
the band, one must take into account the other effects previously
mentioned.

In addition to calculating the PP DRS and DTS signals, we have also
calculated the FWM signal. The calculations agree with the experimental
results of Fig.~\ref{fig2} showing: (i) the decay of the FWM signal as
one increases the detuning, $\Delta$ and excites further into the band,
(ii) that that the EID does not change the rise-times of the FWM at the
exciton energy or in the band, and (iii) that the oscillations in the
FWM for positive time delay are out of phase with the oscillations in
the DRS.  Similar results were also seen in calculations of the  FWM in
high-quality GaAs quantum wells.\cite{Sayed}

\begin{figure}[tbp]
\includegraphics[scale=0.45,trim=0 0 0 10,clip]{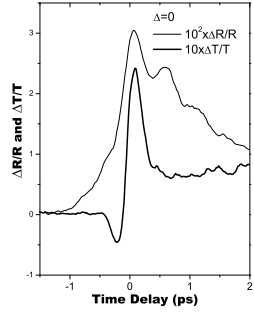}
\caption{Experimental DRS (solid line) and DTS (dotted line) for a 2
$\mu$m-thick GaN sample. The DRS and DTS signals were multiplied by the
values in the figure for comparison purposes and the frequency of the pump was
chosen to be between the A and B excitons ($\Delta=0$). The DRS shows a slow rise time for
negative delays while the DTS does not show a slow rise time similar to the
calculations in Fig.~\ref{fig5}(a).  \label{fig6}}
\end{figure}

To investigate futher our theoretical prediction that the DRS signal
shows a slow rise time while the DTS signal shows a fast rise time,
further experimental pump-probe measurements were performed in both the
reflection and transmission geometry on bulk GaN. To be able to perform
transmission measurements, the 2 $\mu$m thick samples were used. The
results are shown in Fig.~\ref{fig6}.  The excitation spectrum was
chosen to excite both the A and B exciton simultaneously. The temporal
traces of the DRS and DTS show the qualitative behavior similar to our
theoretical calculations including EID; a much slower rise time in the
DRS as opposed to the more rapid, pulse-form limited rise of the DTS.
Note that there is a negative dip in DTS near zero delay.  This can
possibly be caused by distortion of the pulse.\cite{Kim} We note that
the thickness of 2 $\mu$m is much larger than the penetration depth of
GaN, hence, any excitonic signals must have been absorbed and the DTS
signal is an order of magnitude larger than DRS signal due to
relatively small transmission compared to the pump induced transmission
change.

One possible reason for the differences in the early time behavior of
the DTS versus the DRS signals is that they measure different
quantities.  The DTS signal is related to the imaginary part of the
dielectric function which is a local function of frequency  depending
on the central frequency of the probe pulse.  The DRS depends more
strongly on the real part of the dielectric function. Hence, through
the Kramers-Kronig transformation, the DRS is sensitive to spectral
regions above and below the central probe frequency.

\begin{figure}[tbp]
\includegraphics[scale=0.45,trim= 0 60 0 50,clip]{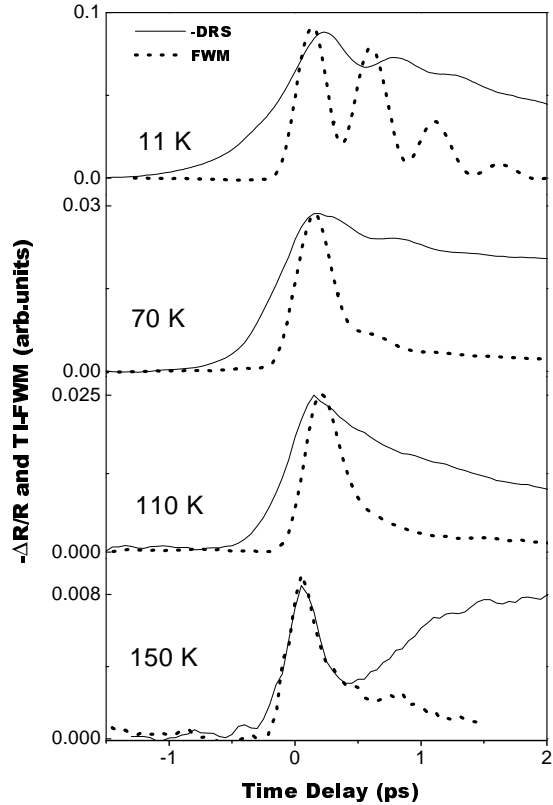}
\caption{Temperature dependence of the FWM (dotted line) and DRS (solid line) signals near
the exciton resonance at 10, 70, 110, and 150 K
for a 7.2 $\mu$m-thick GaN epilayer. The detuning $\Delta = 0$. \label{fig7}}
\end{figure}

In Fig.~\ref{fig7} we investigate the temperature dependence of the DRS
and FWM for $\Delta=0$.  The polarization decay time is given by the
eq.~\ref{scatter}.  The first term on the right
represents decay due to scattering with impurities or phonons which is
not strongly density dependent and the second term represents the
density dependent scattering mechanisms such as carrier-carrier
(electron-electron, electron-hole etc). We expect the first term to be
much more temperature dependent since, for example, electron- phonon
scattering depends on the phonon occupation number which depends
strongly on temperature.  Carrier-carrier scattering is only weakly
temperature dependent.  Fig. ~\ref{fig7} shows the data of the DRS
(solid lines) and TI-FWM (dotted lines) measured at four different
temperatures (T=11, 70, 110, and 150 K). The laser was tuned to excite
both the A exciton and B exciton. We note that as the temperature is
increased, the rise time at negative delay of the DRS becomes
more rapid and  approaches that of the FWM.  We see that near 150K ,
both the DRS and FWM have nearly the same rise time. In addition, the
DRS rise time and FWM decay time (for positive delay) show a similar
tendency to decrease with increasing temperature.  The rapid decrease
of FWM decay time starting from around 150 K is due to the dominance of
optical phonons as a scattering mechanism for dephasing of
excitons.\cite{Fischer} Note that the rapid phonon scattering time above
150 K also has the effect of rapidly damping out the FPD and hence the
DRS signal now has a fast rise time.

Our theoretical results based on EID are suitable only for the low
density regime where the nonlinear response near the band edge is
dominated by excitonic screening of the carrier-carrier Coulomb
potential.\cite{Wang} As we increase the carrier density ten-fold
as shown in Fig. 8, EID is not dominant any more and the
slow rise signal which was observed in Fig. 2 is superimposed by a
faster rise with opposite sign. The sign change in Fig. 8
could possibly be associated with the band-gap renormalization and
reduction of the Coulomb enhancement factor which occurs for high
density photoexcitation.\cite{Sun1,Langot}

\begin{figure}[tbp]
\includegraphics[scale=0.45,trim=0 250 0 0,clip]{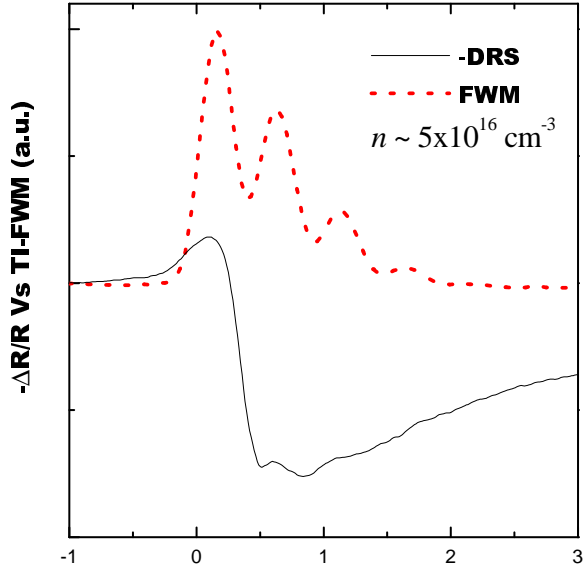}
\caption{The DRS (solid line) and FWM (dotted line)
in a  7.2 $\mu$m-thick GaN
epilayer, for high carrier density photoexcitation ($\sim 5 \times
10^{-16} cm^{-3}$).
\label{fig8}}
\end{figure}

\section{Conclusion}

In this work, we have studied the initial temporal dynamics in GaN
epilayers through the simultaneous measurement of PP DRS and FWM. For
resonant excitation of the A and B excitons, we have observed an
unusually slow rise time in {\it negative time delay} only in the PP
DRS in contrast to the FWM and DTS signals which show a more rapid rise
time.  These differences can be explained by excitation induced
dephasing and the fact that the FPD of the probe pulse can contribute
to the PP signal, but not the FWM.  We have shown  from simulations of
the Semiconductor Bloch Equations, that EID strongly alters the
line-shape for the DRS signal.  With no scattering or excitation
induced dephasing, the negative time delay in the DRS shows
oscillations.  With scattering and EID, we obtain a slow rise time at
the exciton resonance for the DRS but not for the FWM or DTS signal.
For energies above the band edge or at higher temperature where
scattering is much stronger, the DRS signal has a short rise time and
the FWM signal decays rapidly even at positive delay times.

\acknowledgements
This work was supported by MOST (the National
Research Laboratory Program) and KOSEF (the Center for Strongly
Correlated Materials Research, and Grant No. 97-0702-03-01-3). The
work at U. of Florida was supported in part through Department of
Energy through grant DE-FG05-91-ER45462 and the National Science
Foundation through grant DMR-9817828.

\end{document}